\documentclass[a4paper,12pt]{article}
\usepackage{natbib} 
\usepackage[pdftex]{graphicx}
\usepackage{epstopdf}
\pdfoutput=1

\usepackage[pdftex]{hyperref} % should be the latest loaded package

\def\jnt#1#2{#1}

\title{Effect of the hypomagnetic field on the size of the eye pupil} 

\author{
V. N. Binhi, R. M. Sarimov
}

\date{}

\begin{document}
\begin{sloppypar}

\maketitle   

A.M. Prokhorov General Physics Institute of the Russian Academy of Sciences; 
Vavilov St., 38, Moscow, 119991; 
email: binhi@kapella.gpi.ru

\begin{abstract} 

Previously, we reported that the hypomagnetic field obtained by the 100-fold deprivation of the geomagnetic field affected human cognitive processes as estimated in several computer tests. The exposure to the hypomagnetic field caused a statistically significant increase both in the task processing time and in the number of errors. The magnitude of this magnetic effect, averaged over 40 healthy subjects and more than $10^5$ separate trials, was about 1.7{\%}. In the present work, the results of a simultaneous study are described, in which the right eye of each subject was video recorded, while the subject performed the tasks. It has appeared that the pupil size grows in the hypomagnetic field. This effect has been calculated based on the treatment of a large data set of about 6$\times 10^6$ video frames. Averaged all over the frames, the magnetic effect on the pupil area was about 1.6{\%}, with high statistical confidence. This is the first laboratory study in which the number of separate trials has been large enough to obtain rather smooth distribution functions. Thus, the small effect of the hypomagnetic field on humans has become evident and statistically valid.
\\ [2mm]
\end{abstract}

\section{Introduction}

The role of the geomagnetic field in life processes remains unclear. Even the fact that some animals can navigate using the geomagnetic field is not yet explained \citep{Johnsen.ea.2008,Mouritsen.2012}. The nature of biological effects caused by such weak magnetic fields is a physical problem \cite{Binhi.ea.2003.PU}. 

There are both epidemiological \citep {Schuz.ea.2009} and laboratory \citep{Ghione.ea.2004,Cook.ea.2006} studies showing some association between the level of AC electromagnetic fields and human health. However, relatively little is known about the effects of weak static magnetic field, on the order of the geomagnetic field (GMF), in humans. 

There were only a few laboratory studies focused on the cognitive effects of weak static magnetic fields, particularly the hypomagnetic field. 

In \cite{Beischer.1971}, 24 subjects, two people at a time, were continuously exposed to 50~nT hypomagnetic field for up to two weeks. A range of psychological tests were performed before and after the magnetic exposure: the space perception test, visual spatial memory, the hand-eye coordination, the reproduction of time intervals, the subject's equilibrium. In all these tests no significant difference was found between the data collected in the geomagnetic and in the hypomagnetic environments. However in \cite{Thoss.ea.2007}, averaged over 55 subjects, the sensitivity of the human eye to a visual light stimulus in the hypomagnetic field was less than that in the GMF by (6$\div $7){\%}. Thus, data available on the effect of HMF on human cognitive processes were insufficient and inconsistent.  

In our earlier work \cite{Sarimov.ea.2008e}, we reported that deprivation of the GMF to the level lower than 400~nT affected human cognitive processes. Forty people, who all gave their informed concent, were tested in a series of four cognitive tests. Under HMF, both the number of errors and task processing times increased by about (1.5$\div$2.5){\%}, on average. These results were obtained by using several multivariate statistical methods: MANOVA, the discriminant, the factor, and the cluster analyses. The total magnetic effect, calculated as the average over about 120000 trials, was (1.7$\pm $0.2){\%}. This value was rather steady: When the array of data was limited to the measurements of the task processing times only, the average effect was 1.64{\%} \cite{Binhi.2012e}; if the results of six subjects who showed maximal effects were removed from the array, the average effect, then 1.49{\%}, retained its statistical significance at $p< 0.004$ \cite{Binhi.ea.2009.EMBM}. So, within the limits of this study, the global mean magnetic effect in humans was formed by the bulk of the measured data and by all the subjects. The observed magnetic effect was  the consequence neither of particular efficiency of any test used, nor of the presence of particularly sensitive subjects. Temperature and atmospheric pressure were studied among possible essential factors, but they did not affect the results. 

It should be noted that all eight measurable characteristics were subjective psychological reactions. It was interesting to understand as well, whether the hypomagnetic field can influence human reactions that are mostly independent of the will of a subject. The pupil size is a characteristic clearly involved in the execution of the aforementioned psychological tests. Although psychologically induced pupil constriction/dilatation is known, a physiological reaction to light, the pupillary light reflex, is well expressed. For this reason, the pupil size has been chosen for tracking simultaneously with the above testing of the subjects under HMF/GMF exposure.   

The aim of conducting the present study was to investigate whether the hypomagnetic field can cause the eye pupil to change in size.

\section{ Method}

\subsection{Subjects}
No special selection of subjects was made but for equal numbers of men and women and of people aged less than and more than 40 years. There were 20 subjects in each gender-specific or age-specific group, and each subject was tested both in GMF and HMF conditions. 

\subsection{Exposure system}

GMF deprivation has been reached by the compensation of GMF in a special wooden box of the size $1$$\times $$1 $$\times $$1.5$~m$^3$. The box included a wire mesh that shielded a test subject from the outer randomly variable electrostatic field. The magnetic field inside the box was measured by fluxgate sensors fixed near the head of the subject, approximately at the center of the box. A digital feedback system compensated (along the main axis) the outer magnetic field and its variations caused by the city electric vehicles and industrial pulses.

Four circular coils 1~m in diameter were spaced at 0.5~m while having 40 windings in the side coils and 26.5 in the middle ones. The total active electrical resistance was 1.23~Ohm. The MF inhomogeneity inside the workspace of the system did not exceed 2{\%}. The main axis of the system was oriented parallel to the GMF (44~$\mu $T) vector at a precision of 0.5 degree. The bandwidth of the feedback system was about 10~Hz, at the MF measuring rate 1000~Hz. The residual value of MF inside the box during experiments did not exceed 0.4~$\mu $T along the main axis and 0.6~$\mu $T in perpendicular directions.

\subsection{Test procedures}

Each subject has been tested twice; the second session was conducted usually in 30$\div $50 days after the first one. In one of these two sessions, HMF was used, and in the other, for comparison, there were the same conditions but without GMF deprivation. To exclude the possible contribution from the order of HMF and GMF sessions, the order of those for a half of the subjects was opposite to those for the other half. Measurable were the task processing times and the number of errors in the following tests: (i) the rate of a simple motor reflex, (ii) recognition of colored words, (iii) short-term color memory, and (iv) recognition of rotated letters. Two of these were modifications of the well-known J.R. Stroop and R.N. Shepard tests. 

A total of eight parameters were measured in this study. The protocol of this experiment is described in detail in \cite{Sarimov.ea.2008e}. What is essential is the following: each of 80 experiments consisted of three time periods: the 10 min of accommodation to the environment at GMF conditions and preparing to be tested; 10+10 min of testing to collect reference data, also at GMF conditions; and 10+10+10+10 min of testing under GMF conditions (in 40 ``sham'' experiments) or under HMF conditions (in other 40 ``real'' experiments). One-minute relax intervals were placed between all these 10-min periods, so that the total duration of an experiment was 76 min. Test subjects were not aware of which magnetic field, GMF or HMF, they were subjected to during 40 min of ``exposure.'' 

A special device has been made for recording eye movements. The plastic frame that was fixed on the subject's head carried an analog video camera ACE-S560H (0.05 lux, 600 lines). A filter was mounted in front of the camera inside the camera cylinder to cut off light with wavelength less than 810 nm. The sensitivity of the camera was enough to work in the IR range. IR LEDs that were placed around the camera aperture illuminated right eye area, which made it possible to significantly contrast the eye pupil. 

The pupil movements were recorded in a digital format MPEG-4 converted to 8-bit gray by means of a video capture device. The rate was 25 fps; the duration of each of 80 records was 76 min. It can be easily derived that a total of about 9 million frames have been collected in this study, half for GMF-, and half for HMF-type experiments.

An original computer program has been developed that could treat the footages, frame by frame. After 80 records were made,  it turned out that one record failed due to a technical fault. So the results of the corresponding subject were removed from the data set, and the program processed only 78 video files of 39 subjects. A preliminary treatment was as following. 

First, the program cut the fragments of the footages that corresponded to the accommodation/relax intervals and to the short intervals of eye blinking. What was left were a little less than 20 min of the reference interval (control GMF conditions), and 40 min of the sham (GMF) or real (HMF) exposure. So we had 39 one-hour footages of GMF/GMF type, i.e., those of ``sham'' or ``simulation'' experiments, and 39 movies of GMF/HMF type, or movies with ``real'' exposure to HMF. 

\begin{figure}[htbp]
\centering 
\includegraphics[scale=0.5]{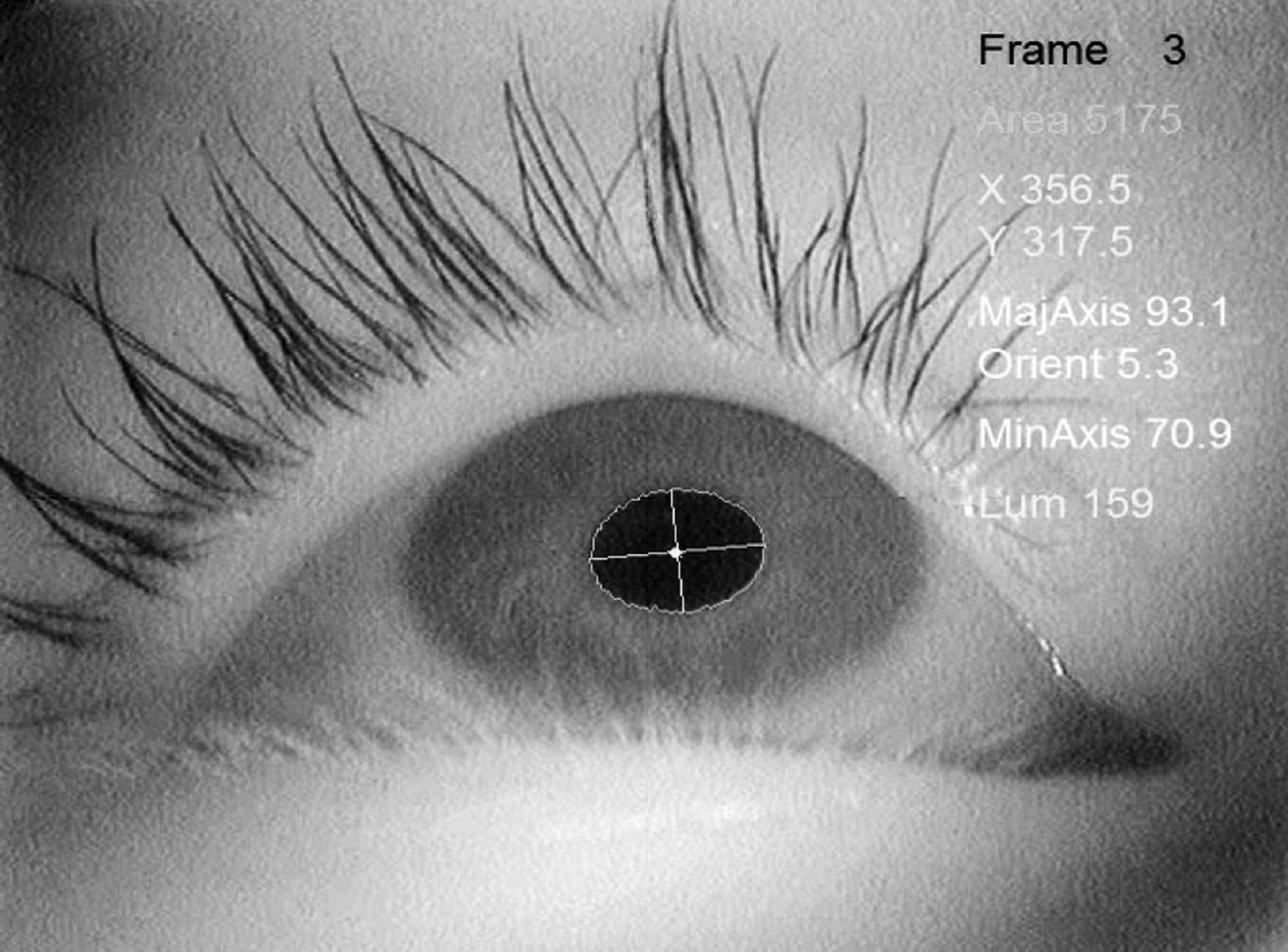} 
\caption{Parameters of the equivalent ellipse of the eye pupil as determined by a computer program. 
}\label{Eye}
\end{figure}

For each frame (680$\times $572~pixel$^2$) of the movies, the program found the image of eye pupil, approximated the pupil by an ellipse, and determined its parameters: short and long axes, rotation angle, horizontal and vertical positions of the pupil, Fig.~\ref{Eye}. These values were saved in a file also containing a timestamp, frame sequence number, and the mean luminance of the frame (the mean density of the gray within 8-bit range 0--255). 
The mean luminance was calculated for the entire frame area except the area of the eye pupil. 

The results of each of 78 experiments were presented as a table/file, each line of which corresponded to a single frame and included the data of its treatment. Each column of the file represented an array: sizes of the ellipse axes, frame luminance, etc. 

The second step of treatment was in examining the written arrays for outliers. Due to the contribution of many uncontrollable factors, the arrays contained not only regular changes, but also a noise. Some values in the arrays can deviate from the means so far that their artifact origin is very probable. Such data are often removed from samples. The program removed an entire frame's line from the file, if one of the values in the line was spaced from the corresponding sample mean by more than three standard deviations. The reduced arrays were used for further calculations.   

While a subject is tested, his or her eye rotates in different directions, so the eye pupil is seen from the camera aperture under different angles, as an ellipse. The actual size of the pupil is closer to the ellipse's major axis, because the minor axis varies as a cosine of the angle of view. We used the major axis as a main observable that was determined for each frame. 
 
In what follows, the arrays of measured pupil sizes, corresponding to the control, or reference, 20-min interval and to  ``exposure'' 40-min intervals, are denoted as $\bf c$ and $\bf x$ for ``real'' experiments, and $\bf s$ and $\bf y$ for ``sham'' experiments, respectively, or, in a different order, $\bf c$ and $\bf s$ stand for controls, and  $\bf x$ and $\bf y$ stand for exposure intervals. Let $c$, $x$, $s$, and $y$ be sample means of those arrays, $\sigma_c$, $\sigma_x$, $\sigma_s$ and $\sigma_y$ be their standard deviations, and $i$ stand for the index that numbers the subjects.

Mathematical operations like ${\bf x}/c$ imply that multiplication by $1/c$ is applied to each element of the array $\bf x$. Then we could determine the result of a subject exposure in ``real'' experiment as the mean of the array ${\bf x}/c$, or a normalized effect $x' \equiv x/c$ that is centered on unity.

However, that would not be a magnetic effect, because the change from $c$ to $x$ could be due to natural physiological rhythms, to learning in the course of testing, etc. Right determination of the magnetic effect of the real HMF exposure would be only in its comparison to the result of ``sham'' experiment, where the mean of the array ${\bf y}/s$, or $y' \equiv y/s$, is calculated. Thus, we determine mean magnetic effect as $m = (x'- y') / y'$. As can be seen, with such determination, the mean magnetic effect can be considered as the mean of the array ${\bf m} = ({\bf x}/c - y')/y'$. This is the array of ``elementary magnetic effects'' defined for each separate frame of the $\bf x$ array. It is convenient, because it makes possible to build different distribution functions and to compare their statistics.  

\section{Results}

First, the magnetic effect has been calculated in average all over the subjects. All the normalized arrays of each subject were combined in single arrays, $\bf X$ and $\bf Y$, separately for ``real'' and ``sham'' experiments:

\[ {\bf X} \equiv \bigcup_i {\bf x}_i /c_i, ~~ {\bf Y} \equiv \bigcup_i {\bf y}_i /s_i . \] 

\begin{figure}[htbp]
\centering 
\includegraphics[scale=0.65]{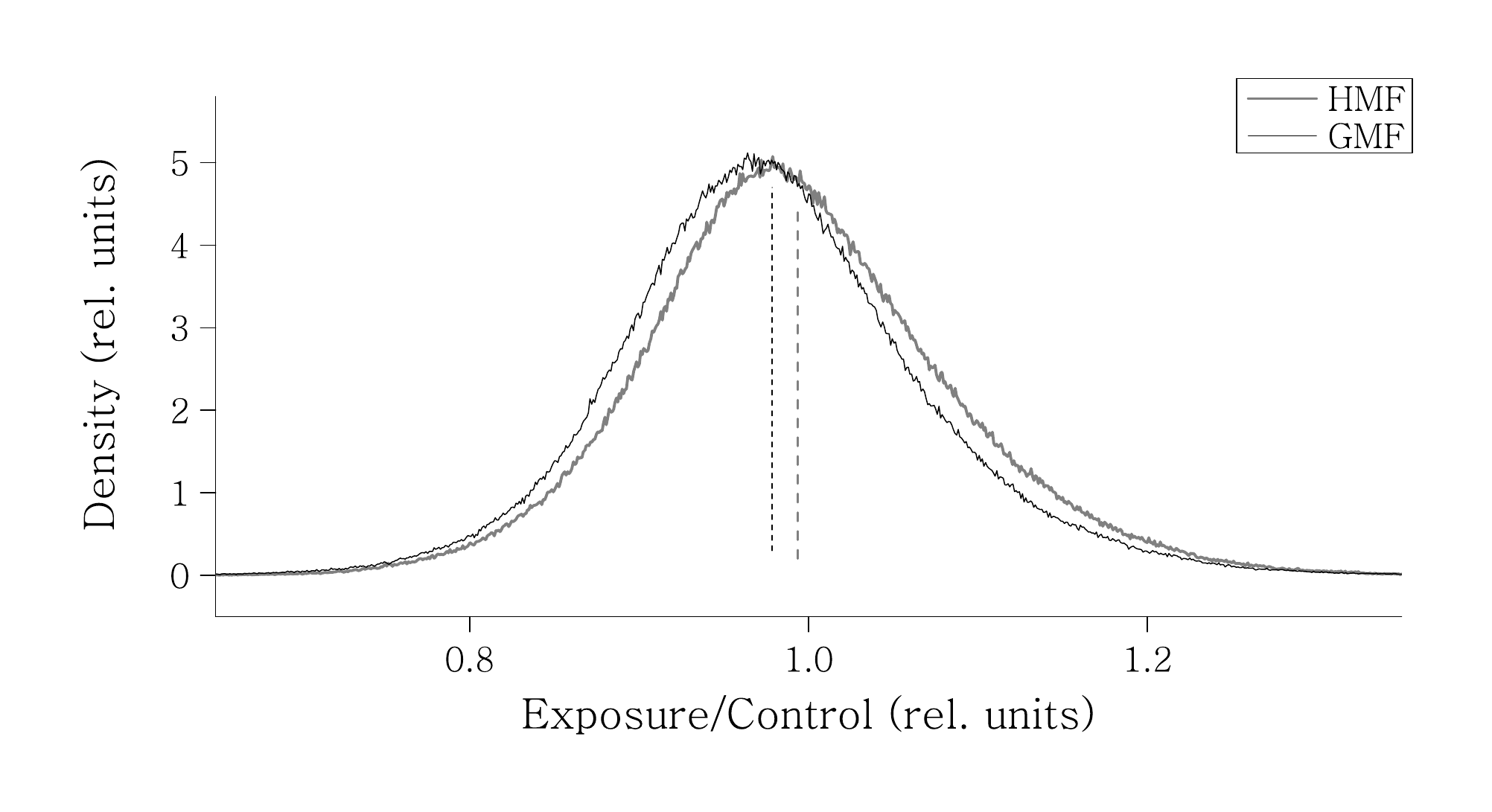} 
\caption{Distributions of the normalized pupil sizes at ``sham'' (GMF) and ``real'' (HMF) exposures, shown as histograms of 1000 bins. The means (dash lines) and the standard deviations of the samples were: $X=0.9935$, $\sigma_X = 0.0908$ and $Y = 0.9785$, $\sigma_Y = 0.0897$ for HMF and GMF correspondingly.  
}\label{HMFvsGMF}
\end{figure}

The distributions of the $\bf X$ and $\bf Y$ elements, i.e., pupil sizes normalized to their means in controls, are shown on Fig.~\ref{HMFvsGMF}. It is the distributions over the relative pupil size values, which are built as histograms, or relative frequencies of corresponding values. The distributions are shown as normalized to unit area under the curves. The arrays' lengths were 1692192 for $\bf X$ and 1671263 for $\bf Y$.

As can be seen, the distributions are different in their means. The distributions are close to the normal one. Two-sample $t$-test shows that the difference is statistically significant with an infinitesimal probability of error ($t$-statistics equals 152). As to the magnitude of the magnetic effect, the ordinary definition gives $M \equiv (X-Y)/Y \approx 1.53${\%}.  

\begin{figure}[htbp]
\centering 
\includegraphics[scale=0.65]{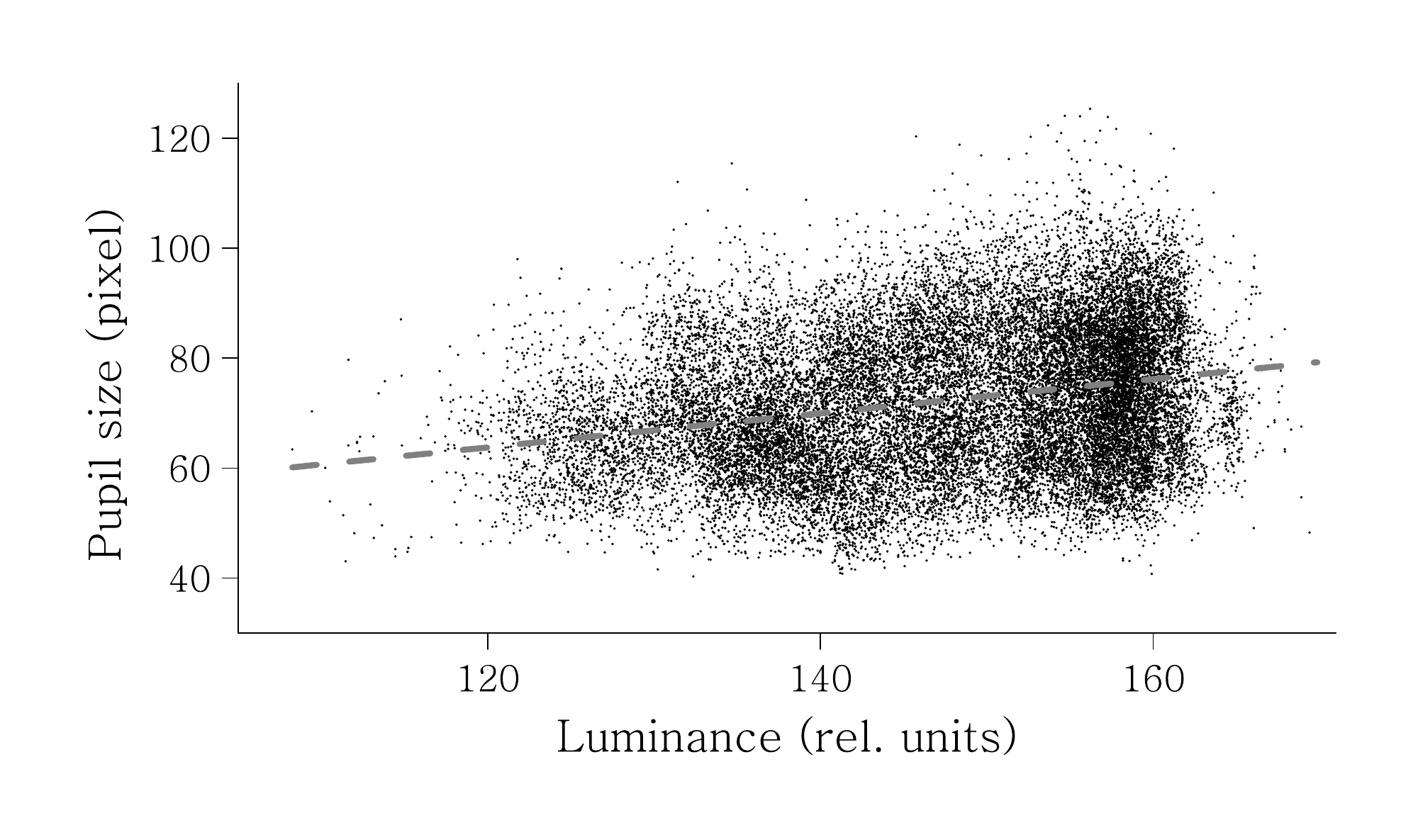} 
\caption{A correlation diagram between the pupil size and the illuminance of the area around the pupil as measured by the frame luminance. Every hundredth point of more than three million is only plotted; the regression line is calculated for the entire set of points.
}\label{correlationLumSize}
\end{figure}

The illumination of the eye in our experiments could possibly vary due to many reasons. It is both the natural and artificial room daylight variations, light from the moving objects on the LCD monitor in front of a test subject, and its individual position in the magnetic exposure box. Despite the effective spectrum of measuring radiation was shifted to the IR range, optical radiation variations could contribute to the outcome because of the pupillary light reflex. Therefore, we paid particular attention to this fact. The mean illuminance of the area around the eye pupil was calculated along with the size of the eye pupil, for each frame. It appeared that there is a direct correlation between these two values, and not an inverse one as could be expected, Fig.~\ref{correlationLumSize}.
  
The reason for this is that the position of the camera was not fixed relative to the face of a subject. A subject could adjust the position in the course of a session so that the distance between the camera and the eye often varied. The smaller the distance, the greater was the illuminance due to the IR LEDs and the greater was the size of the pupil apparent to the camera; it is a geometric constraint. At the same time, the average luminance over all the frames in HMF experiments has occasionally been greater than in GMF experiments. For this reason, we had to suggest that the observed increase in pupil size under HMF was at least partly caused by this geometric effect. Therefore, a correction was necessary to allow for the correlation and exclude the geometric effect of luminance.

The correction procedure was to determine the coefficients of simple linear regression and to correct pupil sizes by subtracting corresponding contributions of the regression. The slope of the regression line in Fig.~\ref{correlationLumSize} is $b=0.3230$, so that corrected values for the sizes have been calculated as $a_{\rm corr} = a - b (E - E_{\rm mean})$, where $a$ is an element of arrays ${\bf x}_i$, ${\bf c}_i$, etc., $E$ is the corresponding luminance, and $E_{\rm mean}$ is the mean luminance averaged over the entire data set. Of course, no correlation was found between the values of the frame luminance and corrected pupil sizes. Nonetheless, the magnetic effect has stood 100{\%} significant ($t$-statistics equals 77, and that figure differs from 100{\%} by a number less than $10^{-1280}$...), at a reduced value though.  

The mean magnitude of the effect from magnetic exposure can only be computed rather than directly measured, so it depends on the definition. A definition $M \equiv (X-Y)/Y$ gives for the corrected set of data $M \approx 0.79${\%}. The pupil size distributions corresponding to the ``real'' and ``sham'' experiments, i.e. those built on the arrays $\bf X$ and $\bf Y$, are practically the same as in Fig.~\ref{HMFvsGMF}, with a smaller gap (not shown). It is essential that if the \textit{area} rather than the size of pupil was used to calculate the magnetic effect, it would be twice as large, $ 1.58${\%}, with also twice the standard deviation of the distributions. With any definition, the magnetic effect is 100{\%} statistically significant. Statistical significance between ``GMF'' and ``HMF'' persists at $p< 10^{-4}$  even if four people, showing maximal positive magnetic effects ($\sim $ 13, 12, 10, and 7{\%}) are removed from the sample. 

\begin{figure}[htbp]
\centering 
\includegraphics[scale=0.6]{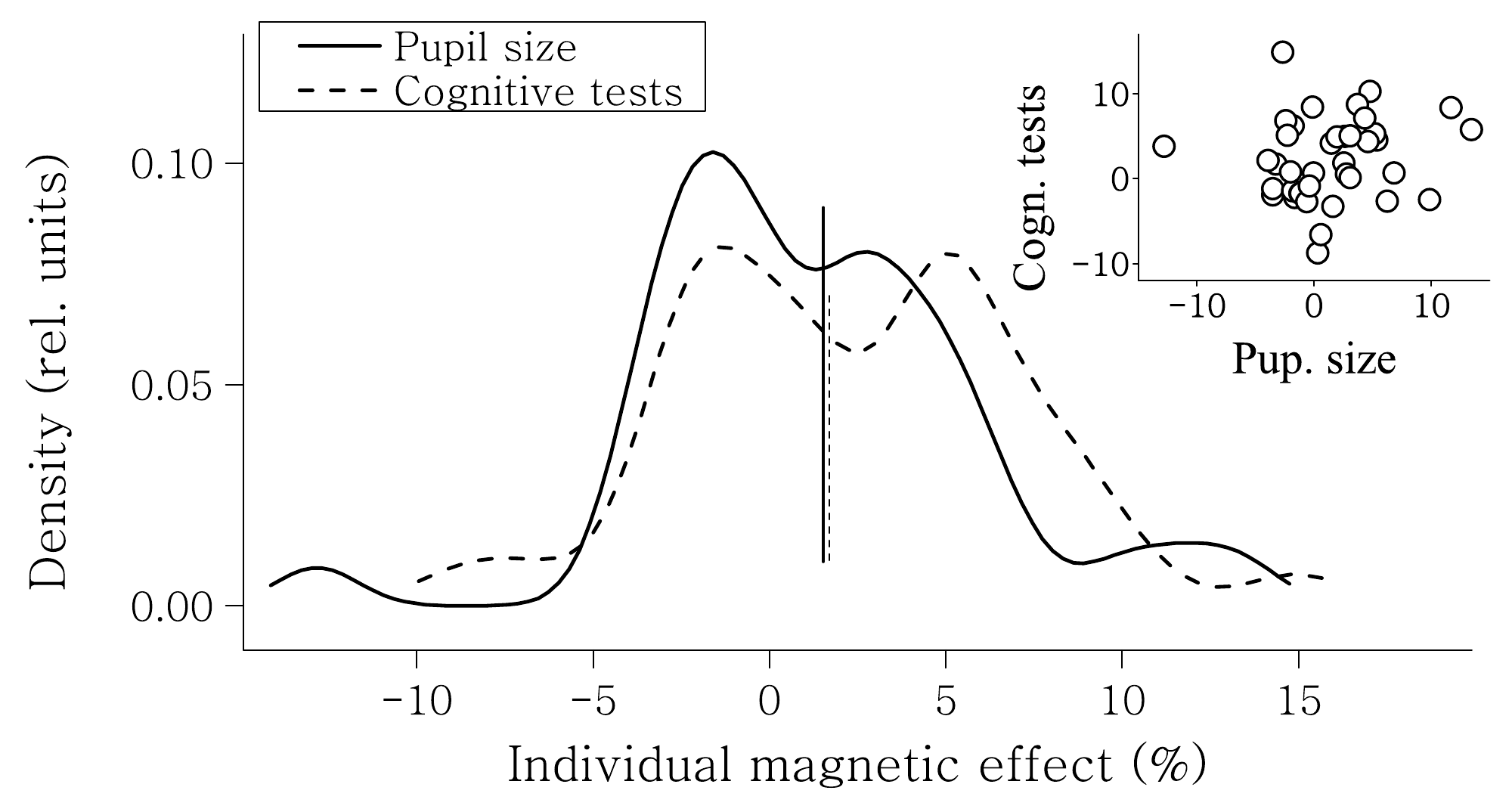} 
\caption{Normalized distributions of the individual magnetic effects calculated for the same subjects: in the present work from the pupil size measurements (solid line, $\sigma = 0.048$) and in the work \cite{Sarimov.ea.2008e} from the parameters of cognitive tests (dash line, $\sigma = 0.061$); mean values are shown. Points in the inlet plot: a correlation diagram for the individual magnetic effects. 
}\label{PupSize-CognTests-Distribs}
\end{figure} 

Individual magnetic effects were studied after the regression correction was made for pupil size in each frame. The individual effects reflect the individual sensitivities of subjects to 40-min HMF exposure. The individual magnetic effects in its ordinary definition have been calculated for each test subject. For this, arrays ${\bf m}_i = ({\bf x}_i/c_i - y'_i)/y'_i$ were separated, and for each one the mean value $m_i$ was calculated. The number of available quantities $m_i$, 39, was enough to compose an array ${\bf u} \equiv \bigcup _i m_i$. The distribution of its elements is shown in Fig.~\ref{PupSize-CognTests-Distribs}. Also shown is the same distribution calculated from the parameters of the cognitive tests (see below). The distributions are given as the density estimation functions with a Gaussian kernel of width equal to 0.25 standard deviations, which corresponds to a histogram of about eight bins in its main interval from $ - \sigma_u $ to $\sigma_u $. 

It is essential that the individual mean magnetic effects, taken separately, were statistically significant. All the subjects except two, who showed the smallest means 0.13{\%} and 0.04{\%}, had their mean magnetic effects significant at the level $p< 10^{-3}$, at least. The ``real'' and ``sham'' distributions for each one of the test subjects are similar to those in Fig.~\ref{HMFvsGMF}, however with greater gaps and noise.

\section{Discussion}

The present study demonstrates that the 40-min exposure to HMF has a statistically significant effect on the subjects: their eye pupils experience a weak dilatation. Despite the total mean effect is small, the distributions of the measured values say something essential about the nature of magnetic effects in human. 

\begin{figure}[htbp]
\centering 
\includegraphics[scale=0.65]{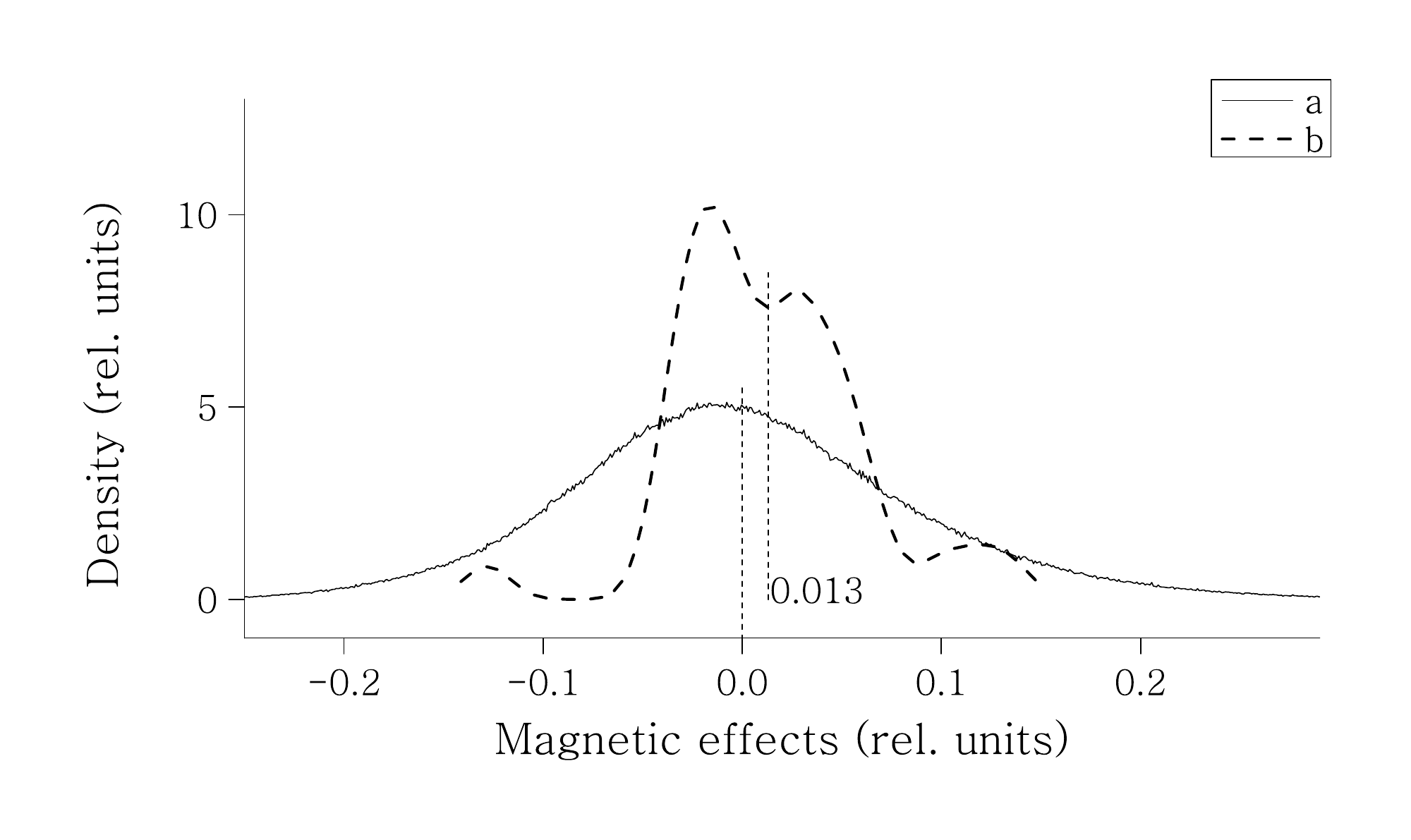} 
\caption{The shape of individual distributions (a) and of the distribution of individual means (b) have essentially different standard deviations, 0.1 and 0.048, respectively. 
}\label{Distrib-ElementaryVSindividual}
\end{figure}

A distribution built on the joint array $ \bigcup_i {\bf m}_i$ mixes two different distributions of magnetic effects that can be isolated. Of interest are the shapes of these distributions. The first one is the \textit{general shape} of individual distributions of the ``elementary magnetic effects,'' i.e., something that is common to individual distributions apart from their mean values. The individual distributions differ by their means, but have something in common---their shape, that can be seen after the means are subtracted from the arrays. In other words, it is the shape of distribution in the joined array ${\bf M} = \bigcup_i ({\bf m}_i -m_i )$, Fig.~\ref{Distrib-ElementaryVSindividual}-a. The other one is the shape of distribution of 39 individual magnetic effects $m_i$ in the array $\bf u$, Fig.~\ref{Distrib-ElementaryVSindividual}-b. The distributions are distinct because the nature of their variances is different. The variance of the first distribution is conditioned by many random factors of brain functioning and physical environment, while the variability of the individual magnetic sensitivity, taken as a biological characteristic, is determined mostly by the phenotypic variation. 

The distributions Fig.~\ref{Distrib-ElementaryVSindividual} show that the mean magnetic effect is not due to the presence of a small hypersensitive group of subjects. Practically all the persons have demonstrated sensitivity to HMF. However, nearly equal parts of the subjects gave opposite responses to HMF that resulted in a small average effect. At the same time, individual magnetic effects varied significantly within the range $\pm (10$$\div$$ 12)${\%}. Thus, the standard deviation is essentially greater than the mean of the individual means. For this reason, the total mean is of little value. It resembles a mean fingerprint pattern, which is actually no pattern at all.

As one can see, a random error in these experiments is very small due to the huge volume of the data set; standard errors of the means are about $7\times 10^{-5}$, which certifies the second significant decimal digit in mean magnetic effect magnitudes. Possible systematic \textit{a posteriori} bias has only been related to the enhanced level of eye luminance in the ``real'' set of experiments. However it has appeared to be inessential, because the pupillary reflex is appreciable only at visible light and not at IR radiation. Apart from the geometric effect, no other possible effects on pupil size have been found. Neither LEDs highlight, nor the artificial indoor lighting, nor the outdoor daylight variations influenced the pupil size, which has been established in a special session of testing. 

As was said above, there were $N=8$ measurands in the cognitive tests. For each of them, the array of the individual mean magnetic effects was separated, and all the arrays were sorted according to the subject's sequence number. Let ${\bf u}^{(n)}$ denote these ordered arrays, where index $n=1,2,..,N$,  is the sequence number of a ``psychological'' measurand used; $n=0$ standing for the measurand of the ``eye pupil size''. Then, one could estimate the correlation between these arrays. A large correlation would signify that one and the same subject possesses higher or lower magnetic sensitivity in different tests, i.e., independently of the measurand used to determine his or her sensitivity. It has turned out that all these arrays \textit{do not correlate}; the mean level of the matrix of correlation coefficients was  

\begin{equation} \label{correlation}  \frac {1} {N(N-1)}  \sum_{n' \neq n}  {\rm corr} ({\bf u}^{(n')},{\bf u}^{(n)})  \approx 0.09 .\end{equation}
This unambiguously confirms that there were no particularly sensitive subjects among 39 tested, although in every separate test there were people showing rather clear response to the hypomagnetic exposure. 

Constrictions and dilatations of the eye pupil occur independently of a human will. It is an objective physiological reaction rather than a reaction based on a subjective will. It is interesting therefore that there is a similarity between the distributions built on both the reactions, Fig.\,\ref{PupSize-CognTests-Distribs}. Essential conclusions follow the fact that this similarity exists together with no correlation between individual means of different measurands.

(1) ``Wings'' are seen in the shape of distributions, at greater absolute values of the magnetic effect magnitudes. The wings, a few percent in area, are not as clear as the main peaks; however they can be seen in the shape of individual distributions both for pupil sizes and for the parameters of psychological reactions. This makes it possible to question the statement that there exist in human population a group of people who are particularly sensitive to electromagnetic fields. It is the so-called ``electromagnetic hypersensitivity syndrome'' repeatedly reported in literature \cite{Schuz.ea.2006}. It states that a few percent of people can markedly react even to relatively weak electromagnetic fields that are incapable of appreciable tissue heating. 

On the face of it, the wing-shaped distribution of individual means does not contradict the hypothesis of hypersensitivity. However, the fact that there is no correlation between the magnetic effects as measured by eye pupil tracking and by psychological reactions, Fig.~\ref{PupSize-CognTests-Distribs} inlet, indicates that people demonstrating a very clear magnetic effect can be different. According to our results, some people tested for a particular biological parameter will clearly react to EMF exposure. However, if a different parameter was chosen to be measured, another minor group would react to the same EMF. We suggest that EMF hypersensitivity exists only as a casual reaction.  

(2) As was said above, individual magnetic effects $m_i$ have been determined for the same subjects, but from their different  characteristics, from the eye pupil size in the present study, on the one hand, and from the number of errors and the test processing time in \cite{Sarimov.ea.2008e}, on the other hand. These magnetic effects have appeared to be uncorrelated. At the same time, the distributions of these effects are rather similar: both have two major peaks and two minor peaks, or wings, in Fig.\,\ref{PupSize-CognTests-Distribs}. This fact indicates that the reaction of a human to MF exposure is not a systemic reaction. 

An external factor, like acoustic noise or light, can cause only a systemic reaction that is conditioned by the human perception, by the functioning of the central nervous system. In this case, different organism's reaction to the external factor should be correlated. Apparently, the same is valid with regard to an internal, but \textit{ab initio} already systemic factor like   a biological rhythm. Unlike such factors of a systemic action, MF is an agent that bypasses human signaling systems, acts directly on tissues, and consequently acts without system, at random. It is exactly this that is observed as the absence of correlation, see (\ref{correlation}), between different biological measurands when a subject is exposed to a HMF. A subject during testing can be magnetically sensitive as measured by one parameter and simultaneously insensitive as measured by another one.  

(3) The data \cite{Thoss.ea.2007} are in accordance with our results \cite{Sarimov.ea.2008e} that changes between GMF and HMF cause a measurable biological reaction in humans. The authors of the former study have concluded that their data are in agreement with the so called ``radical-pair mechanism'', see for example \cite{Gegear.ea.2010}. According to this concept, some animal species have a magnetic sense, because the GMF affects spin-correlated pairs in cryptochrome photoreceptors in the eye retina. The findings of the present study are at variance with this hypothesis. The absence of the correlation between different measurands in (\ref{correlation}) proves that human reaction to magnetic field is not a systemic reaction. Consequently, it is not a reaction caused by the visual analyzer, and in particular, by the changes in its retinal cryptochromes. 

Our data are in a better agreement with the idea that the targets of MF are more or less evenly spread over human organism. It might be magnetic nanoparticles found in human brain tissues \cite{Kirschvink.ea.1992}. Magnetic nanoparticles are small magnets that behave like a compass needle; they can rotate in an external MF. Magnetic nanoparticles produce their own relatively large mT-level MF. In turn, this  MF can affect magnetosensitive radical-pair biochemical reactions \cite{Binhi.2008.IJRB}, so that external MFs as weak as 200~nT can cause biological effects \cite{Binhi.2006.BEM}.   

\section{Conclusions}

The hypomagnetic field of about 400 nT widens the area of the human eye pupil by about $1.6${\%} on total average, with a high statistical confidence. This result is based on human eye video recording at cognitive testing of 39 people in usual geomagnetic environment and under exposure to the hypomagnetic field.

There are two types of the distributions of magnetic effects: (i) averaged individual distribution of elementary magnetic effects and (ii) distribution of the individual mean magnetic effects.

The standard deviation of the distributions is much greater than the mean effects, which makes the total mean magnetic effect uninformative about the unknown nature of magnetic effects in humans.

The distribution of individual mean magnetic effects has a multi-peak shape nearly similar for all tested measurands. For each measurand, the peaks of the distribution are formed by different people. 

The hypomagnetic effect observed in 39 test subjects as measured by eye pupil size and by eight cognitive parameters is likely a general magnetic effect in the human population. Due to the fact that magnetic reactions observed simultaneously with respect to different measurands do not correlate, these reactions to magnetic fields are mostly casual reactions. It takes a large volume of observations in order to register a very weak total magnetic effect.

\end{sloppypar}
\end{document}